\documentclass[aps,prb,notitlepage,superscriptaddress,longbibliography]{revtex4-1}
\usepackage{epsfig}
\usepackage{graphicx}  
\usepackage{dcolumn}   
\usepackage{bm}        
\usepackage{amsmath}
\usepackage{amsfonts}
\usepackage{mathrsfs}
\usepackage{bbm}
\usepackage[caption=false]{subfig}
\usepackage{enumerate}
\usepackage{tikz}
\usepgflibrary{arrows.meta}
\usepackage{physics}
\usepackage{multirow}

\newcommand{\He}{\ensuremath{^{3}\text{He}}}

\newcommand{\upt}{\ensuremath{\text{UPt}_3}}
\newcommand{\ube}{\ensuremath{\text{U}_{1-x}\text{Th}_x\text{Be}_{13}}}
\newcommand{\pos}{\ensuremath{\text{PrOs}_4\text{Sb}_{12}}}

\begin{document}
\title{Generalized Spin Fluctuation Feedback in Correlated Fermion Superconductors}
\author{Adil Amin}
\email{adilamin@uwm.edu}
\affiliation{Department of Physics, University of Wisconsin--Milwaukee, Milwaukee, Wisconsin 53201, USA}

\author{D.F.~Agterberg}
\email{agterberg@uwm.edu}
\affiliation{Department of Physics, University of Wisconsin--Milwaukee, Milwaukee, Wisconsin 53201, USA}

\date{\today}

\begin{abstract}

Experiments reveal that the  superconductors  $\upt$, $\pos$ and $\ube$ undergo two 
superconducting transitions in the absence of an applied 
 magnetic field. The prevalence of these multiple transitions suggests a common 
 underlying mechanism. A natural candidate theory which accounts for these two 
 transitions  is the existence of a small symmetry breaking field, however such a 
 field has not been observed in $\pos$ or $\ube$ and has been called into question for 
 $\upt$. Motivated by arguments originally developed for superfluid $\He$ we propose 
 that a generalized spin fluctuation  feedback effect is responsible for these two 
 transitions.  We first develop a phenomenological theory for $\He$ that 
 couples spin fluctuations to superfluidity, which correctly predicts that a high 
 temperature broken time-reversal superfluid $\He$ phase can emerge as a consequence. 
 The transition at lower temperatures into a time-reversal invariant superfluid phase 
 must then be first order by symmetry arguments. We then apply this phenomenological 
 approach to the three superconductors $\upt$, $\pos$ and $\ube$ revealing that this 
 naturally leads to a high-temperature time-reversal invariant nematic superconducting 
 phase, which can be followed by a second order phase transition into a broken time-reversal symmetry phase, as observed. 
\end{abstract}

\maketitle

\label{sec:intro}
There has been  renewed interest into unconventional superconductors, as
they  provide a natural platform for topological states 
\cite{qi2011topological,sato2017topological,schnyder2008classification,read2000paired}. Correlated fermion superconductors, such 
as $\upt$, $\pos$, $\ube$, URu$_2$Si$_2$ have been intensely studied  as they show time-reversal symmetry-breaking (TRSB)
 \cite{schemm2014observation,levenson2018polar,2015PhRvB..91n0506S,2007PhRvL..99k6402K,heffner1990new}, 
and may host  Majorana modes as well as Bogoliubov Fermi surfaces 
\cite{yanase2017mobius,kozii_three-dimensional_2016,goswami2013topological,agterberg2017bogoliubov}. Of these, $\upt$, $\pos$, and  $\ube$ show a rich 
phase diagram, with two superconducting phases under zero field.  The high 
temperature A phase is  time-reversal symmetric (TRS)  and the low 
temperature B phase is a broken time-reversal symmetry state \cite{luke1993muon,joynt2002superconducting,1985PhRvB..31.1651O,heffner1990new,zieve2004pressure, aoki2003time, adenwalla1990phase}. 

 The presence of two transitions in three different materials raises a question about the underlying mechanism. $\upt$  has been the most studied of 
these materials and has a phase diagram as shown in Fig. \ref{fig2}  \cite{joynt2002superconducting}. The most common 
 explanation for this phase 
 diagram relies  on coupling  the
 superconducting order parameter to a weak symmetry breaking field 
 (SBF), which  splits the degeneracy between the different 
 order parameter components
 \cite{hess1989broken,machida1989unconventional}. The  
  symmetry breaking field 
 is associated with an antiferromagnetic (AFM) order seen in older neutron 
 scattering measurements
 \cite{1988PhRvL..60..615A}. However, recent experiments show that 
 there is no static order near $T_c$, 
 though  antiferromagnetic fluctuations are present \cite{koike1998long,gannon2017spin}, which casts serious doubts about  the use 
 of a symmetry breaking field to generate 
 two transitions. Meanwhile, there is  no accepted model which accounts for the double 
  transition in $\ube$ or $\pos$, though 
there  are signatures of antiferro-quadrupolar 
 (AFQ) fluctuations in $\pos$ and antiferromagnetic fluctuations in $\ube$ 
    as seen in  inelastic neutron scattering (INS) \cite{kuwahara2005direct,hiess2002neutron}. It is natural to ask  if these fluctuations can account for the generic observation of two transitions. 
   
 To gain insight into this question, it is reasonable to consider superfluid $\He$,  which also exhibits  multiple phases. In this case, conversely,  there is a high temperature, high pressure time-reversal symmetry-breaking (TRSB)
 A phase and a low temperature low pressure time-reversal symmetric B 
 phase as shown in Fig. \ref{fig:1} 
 \cite{leggett1975theoretical,vollhardt2013superfluid}. 
 Originally, the stability of the A phase 
 was  a puzzle, as weak coupling theory  predicted that the B 
 state was stable for all 
 temperatures \cite{balian1963superconductivity}. This paradox was 
 resolved by Anderson and Brinkmann, who showed that coupling superfluidity to paramagnetic fluctuations, can stabilize the A state, through a mechanism called the  spin fluctuation feedback effect (SFFE)
 \cite{anderson1973anisotropic, brinkman1974spin, 
kuroda_paramagnon_1975}.

In this paper, we propose a mechanism for multiple transitions 
in {correlated fermion materials by coupling 
superconductivity to fluctuations (both antiferromagnetic and antiferro-quadrupolar), 
analogous to superfluid $\He$. We initially 
formulate a simple  phenomenological method to capture the essential physics of superfluid $\He$, and show that it reproduces the microscopic  spin fluctuation feedback effect  developed by Anderson-Brinkman.     
We then apply this  to $\upt$, $\ube$ and $\pos$  and show that these fluctuations 
change the coefficients of the Ginzburg-Landau  theory and allow for the possible stabilization of a time-reversal symmetric  A phase.  We then consider a transition into the broken time-reversal symmetric state, implementing the symmetry constraints associated with observing a polar Kerr effect, when applicable. These considerations  constrain the possible order parameters. We obtain the following results:

1. We show that except for the 3D $T_{g/u}$ irreducible  representations (reps) of $\pos$,  the only possible way to undergo  two successive transitions,  is for  the B state to be a  time-reversal broken state.

2. We also find that the Kerr effect measurement rules out the 2D $E_{g/u}$ rep scenario for $\pos$. We suggest subsequent Kerr  measurements with different training fields directions may further constrain the order parameters for the  3D $T_{g/u}$ rep case.

3. For $\ube$ in the case of it's 3D $T_{g/u}$ reps, the form of the  spin fluctuation feedback effect  allows for only one A states symmetry out of two  possible states.

4. We also suggest a polar Kerr measurement be conducted  on $\ube
 $, as 
 the presence of a polar Kerr signal would rule out the 2D (E$_{g/u}$) scenario for $
 \ube$, and would constrain the pairing channels for the 3D (T$_{g/u}$) rep scenario of $\ube$.

These results are tabulated in Table \ref{tab:table1}. We have only written one representative state when there are degeneracies,   the other degenerate states appear in the main  text and the appendix.

\begin{figure}
    \centering
    \begin{minipage}{0.45\textwidth}
        \centering
        \includegraphics[scale=0.09]{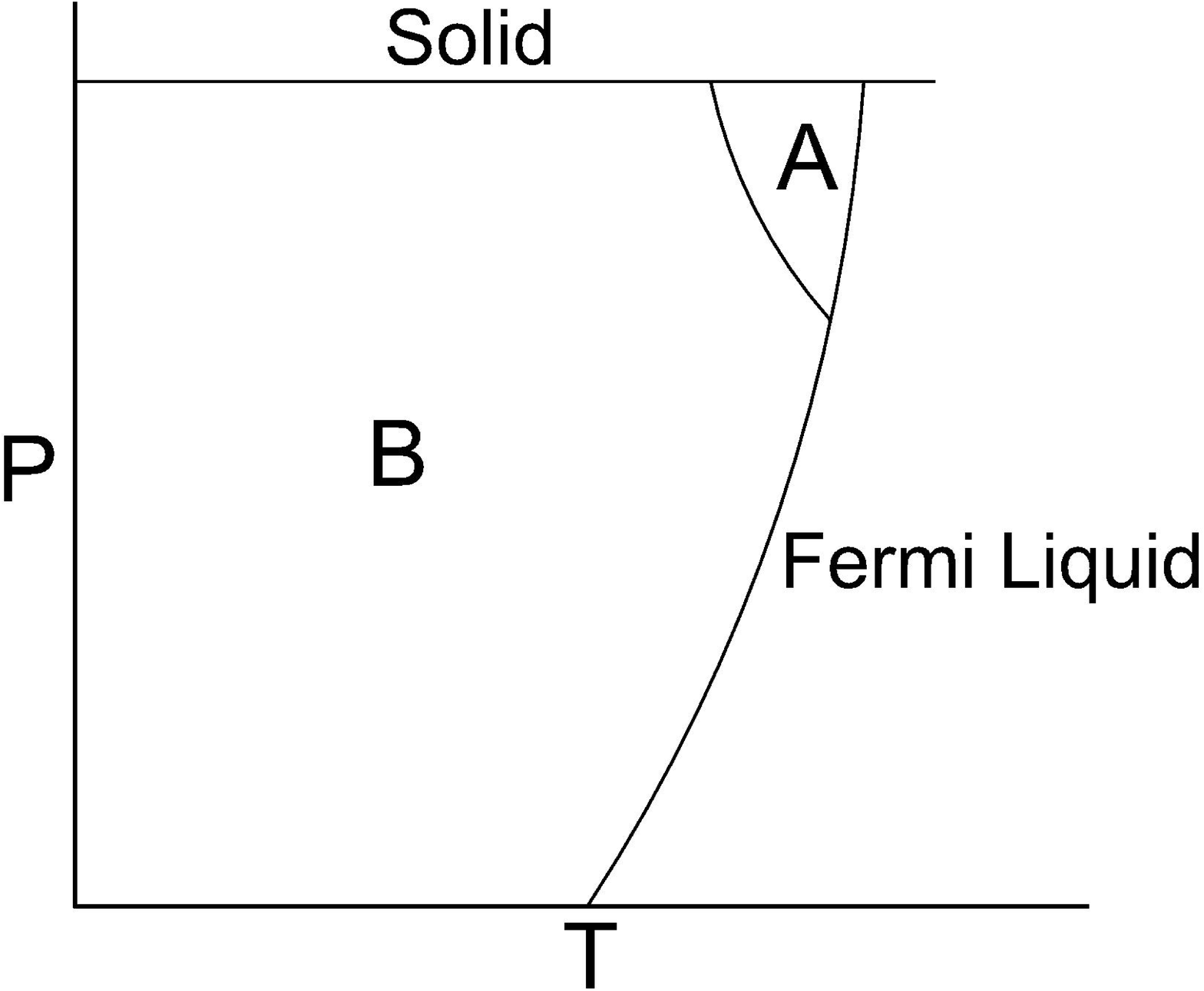}
    		\caption{Pressure (P) - Temperature (T) Phase Diagram of $\He$ }
    		\label{fig:1}
    \end{minipage}\hfill
    \begin{minipage}{0.45\textwidth}
        \centering
    	\includegraphics[scale=0.09]{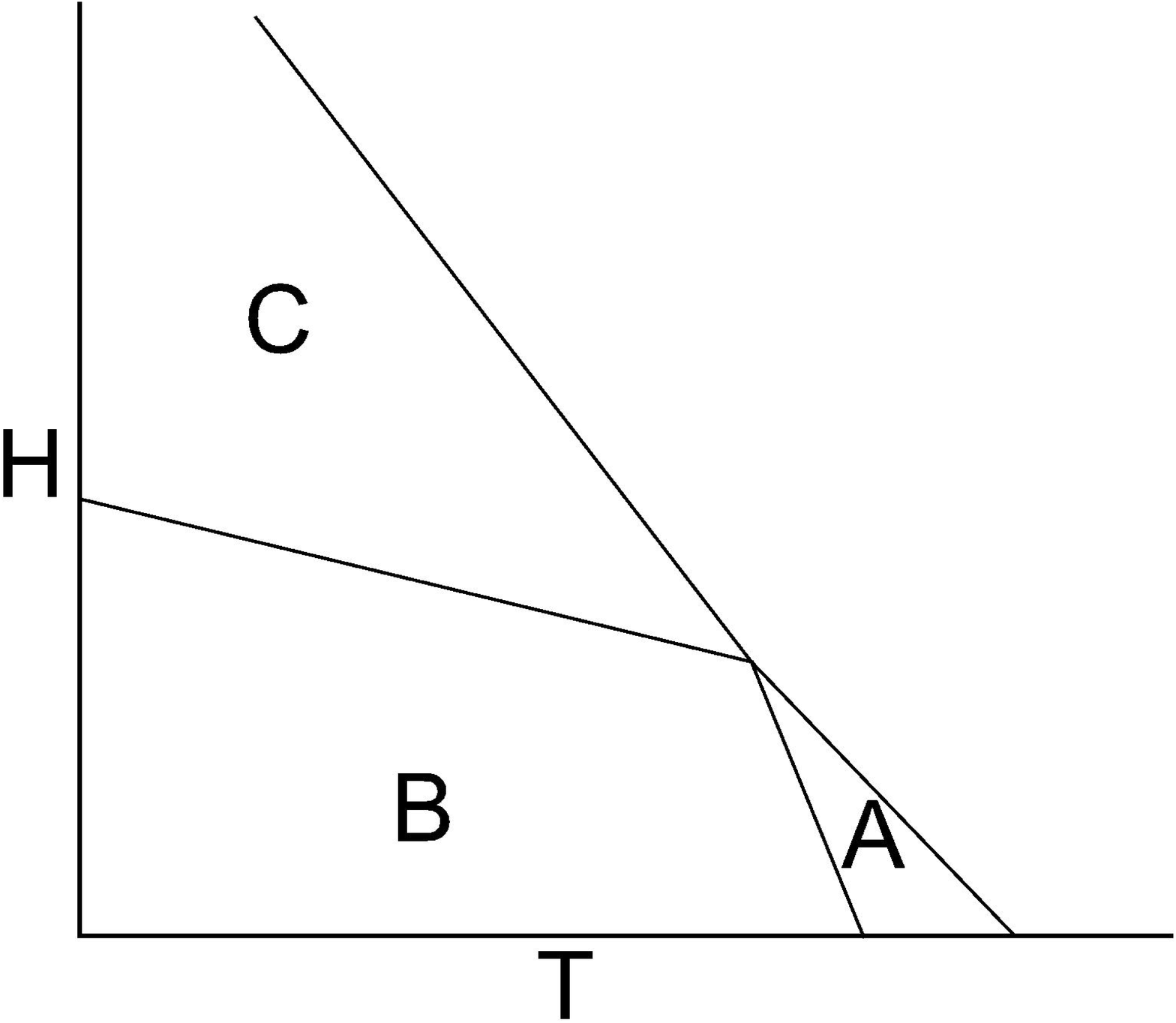}
	\caption{Magnetic Field (H) - Temperature (T) Phase Diagram of $\upt$ with B field $\perp$ z axis}
	\label{fig2}
    	\end{minipage}
\end{figure}

\begin{table}[ht]
  \begin{center}
   
    \begin{tabular}{c c c c c c}

     \textbf{Material} & \textbf{Fluctuations}&\textbf{Reps} & \textbf{Order Parameter} & \textbf{A} & \textbf{B}\\[0.5ex]
     \hline\hline
      
     {$\upt$} & AFM & E & $k_xk_z, k_yk_z$ &(1,0)  & (1,$i\delta$)\\ 
      \hline
      
 	\multirow{3}{*}{$\pos$} & \multirow{3}{*}{AFQ}&{E}&{$2k_z^2-k_x^2-k_y,k_x^2-k_y^2$}& $\times$ &$\times$ \\
 	& & \multirow{2}{*}{T}&\multirow{2}{*}{$k_yk_z,k_xk_z,k_xk_y$} & (1,0,0)& $ (1,i\delta,0)$ or $(1,0,i\delta)$\\
       & & & &(1,1,1) &$(1,1+i\delta\cos\theta,1+i\delta\sin\theta)$ \\
       \hline
       
	\multirow{4}{*}{$\ube$} & \multirow{4}{*}{AFM}&\multirow{2}{*}{E} &\multirow{2}{*}{$2k_z^2-k_x^2-k_y,k_x^2-k_y^2$} &(1,0) & $(1+i\delta,0)^*$ or $(1,i\delta)^*$\\
	& & & &(0,1)& $(0,1+i\delta)^*$ or $(i\delta,1)^*$ \\
	& &T&  {$k_yk_z,k_xk_z,k_xk_y$}& (1,0,0)& $(1+i\delta,0,0)^*$ or $(0,1+i\delta,0)$ \\
	
	\hline \hline
    \end{tabular}
    \caption{Table summarizing our results. The \textbf{fluctuations} column shows the fluctuations  responsible for  spin fluctuation feedback effect, whose Q vector dependence can be found in the text. The \textbf{reps} label the  irreducible representation (reps) of the order parameter. The \textbf{order parameter} gives a representative basis function of the superconducting order parameter and displays the symmetry properties of it's components. The \textbf{A} column   is the time-reversal symmetric state, where each state is multiplied by an overall $\eta_0$ magnitude. The \textbf{B} state is the time-reversal symmetry-breaking B state grown out the A state, where the ``or" separates different  B reps, and we only write one state for each rep, the other degenerate states are in the text and appendix. Here $\delta$ represents a small number, which grows positively from 0 at the second transition. The $()^*$ is for A--B transitions which are  Kerr inactive. The $\times$ represents that this channel is ruled out.}
     \label{tab:table1}
  \end{center}
\end{table}

\label{He3}
\section{\He} 
$\He$ is a strongly correlated Landau-Fermi liquid,  whose quasiparticle excitations  pair to form  a spin-triplet p-wave superfluid 
\cite{leggett1975theoretical,vollhardt2013superfluid}.  The gap function is $\Delta$(\textbf{k}) = $i(d_{i}(\textbf{k})\sigma_i)\sigma_y$, with $d_i=d_{i\alpha} \vu{k}_{\alpha}$, in this manuscript we use the Einstein summation convention.  The order parameter $d_{i\alpha}$ is   
a 3 $\times$ 3 matrix with complex entries, where $i$  is the spin index and  $\alpha$ is the orbital 
index and both run over $x$, $y$ and $z$.  By comparing to experiments, the  $\He$-A phase was  identified  with the  Anderson-Brinkman-Morel (ABM) state with $d_{xx}=\frac{\Delta}{\sqrt{2}},d_{xy}=i\frac{\Delta}{\sqrt{2}}$ and all other $d_{ij}=0$,  while  B state was associated with Balian-Wethamer (BW) which has $d_{ij}=\frac{\Delta}{\sqrt{3}}\delta_{i,j}$
\cite{vollhardt2013superfluid,leggett1975theoretical}. Weak coupling  theory showed that the BW state is stable for all temperatures 
\cite{balian1963superconductivity}, implying that a strong coupling approach was needed to 
explain the existence of the high temperature high pressure A phase.  Anderson-Brinkman \cite{anderson1973anisotropic} used  spin fluctuation feedback effect   to stabilize the A phase, which relied on  the pairing glue in $\He$ being paramagnetic fluctuations. This implies that the formation of the superfluid  alters the pairing interaction, where the type of modification 
depends on which state is formed \cite{leggett1975theoretical}.  Thus the A state can be stabilized despite  being unstable under  weak coupling theory. The A--B transition in this case is first order as the B  state is not a subgroup of the A state.

 Here  spin fluctuation feedback effect  shall be recaptured in a phenomenological manner, by coupling the superfluid 
order parameter to paramagnetic fluctuations, and calculating the resulting change to the bare free energy. 
The bare free energy density  of superfluid $\He$ is given as
\begin{align}
\beta f_{sf}=\alpha d_{i\alpha}d^*_{i\alpha}+\beta_1 d_{i\alpha}d_{i\alpha}d^*_{j\beta}d^*_{j\beta}+\beta_2d_{i\alpha}d_{j\alpha}d^*_{i\beta}d^*_{j\beta}+\beta_3d_{j\alpha}d_{j\beta}d^*_{i\alpha}d^*_{i\beta}+\beta_4d_{i\alpha}d^*_{i\alpha}d_{j\beta}d^*_{j\beta}+\beta_5d^*_{i\alpha}d_{j\alpha}d_{i\beta}d^*_{j\beta}
\end{align}
To this we add the coupling of superfluidity and magnetic fluctuations, which is constructed  to be invariant under independent rotations  in orbital and spin space. The free energy density is given as
\begin{align}
\beta f_{sf-m}= A_1 m_im_i+K_1 
m_im_id_{j\alpha}d^*_{j\alpha}+K_2m_im_jd_{i\alpha}d^*_{j\alpha} +K_3 im_i(\epsilon_{ijk}d_jd_k^*) + B (\nabla_i m_j)^2,
\end{align}

where $m_i$ is the magnetic 
order parameter associated with spin fluctuations. The $K$'s are the couplings between the spin fluctuations and the superfluid order parameter, and the B term is the spatial variation (i.e. q dependence) of the spin order.   We assume  $A_1$ is parametrically  small  and positive (i.e. $A_1 \rightarrow 0$) to indicate that we have large fluctuations. We have the following Hamiltonian
\begin{align}
 \beta \mathcal{H} = \beta \mathcal{H}_0 + \mathcal{K} = \int d^3x \left[A_1m_j^2 + B(\nabla_i m_j)^2\right] + \int d^3x \left[K_1(m_i^2)(d_{j\alpha}d^*_{j\alpha})+\right.
 \left. K_2 m_im_jd_{i\alpha}d^*_{j\alpha} + K_3 im_i(\epsilon_{ijk}d_{j\alpha}d_{k\alpha}^*)  \right],
 \end{align}

 here $\beta\mathcal{H}_0$ is the Guassian theory for spin fluctuations, while $\mathcal{K}$ is the coupling between superfluid order and  the spin order fluctuations. After Fourier transforming we get the following
\begin{align}
 \beta \mathcal{H} = \int \frac{d^{3}\textbf{q}}{(2\pi)^3}\left(A_1 + Bq^2\right)\abs{\tilde{m}(q)}^2 + \mathcal{K},
 \end{align}
 The coupling between superfluidity and magnetic fluctuations  in Fourier space is 
 \begin{align}
 \mathcal{K} &= \int \frac{d^3\mathbf{q}_1d^3\mathbf{q}_2d^3\mathbf{q}_3d^3\mathbf{q}_4}{(2\pi^3)^4} (2\pi)^3 \delta^3(\mathbf{q_1}+\mathbf{q_2}+\mathbf{q_3}+\mathbf{q_4}) \left[K_1\tilde{m}_i(\mathbf{q_1})\tilde{m}_i(\mathbf{q_2})\tilde{d}_{j\alpha}(\mathbf{q_3})\tilde{d}_{j\alpha}^*(\mathbf{q_4})\nonumber\right.\\
 &+K_2\left.\tilde{m}_i(\mathbf{q_1})\tilde{m}_j(\mathbf{q_2})\tilde{d}_{i\alpha}(\mathbf{q_3})\tilde{d}_{j\alpha}^*(\mathbf{q_4})\right]+K_3\int \frac{d^3\mathbf{q}_1d^3\mathbf{q}_2d^3\mathbf{q}_3d^3\mathbf{q}_4}{(2\pi^3)^3}(2\pi)^3 \delta^3(\mathbf{q_1}+\mathbf{q_2}+\mathbf{q_3})i\tilde{m}_i(\mathbf{q_1})\epsilon_{ijk}\tilde{d}_{j\alpha}(\mathbf{q_2})\tilde{d}_{k\alpha}^*(\mathbf{q_3})
 \end{align}

Near the  superfluid transitions, the coupling $\mathcal{K}$ can be treated perurbatively around the Gaussian  theory of the magnetic fluctuations (i.e. $\beta\mathcal{H}_0$). This quadratic theory for spin fluctuations is valid  as long as we are more than a Ginzburg temperature away from the critical temperature associated with magnetic ordering. Thus  the coupling can be evaluated as  
 \begin{align}
 \beta\mathcal{H} = \beta\mathcal{H}_0 - \ln\expval{\exp^{-\mathcal{K}}}_m,  
 \end{align} 

 here $\expval{}_m$ implies  that we are calculating the expectation values with respect to the Gaussian $\beta \mathcal{H}_0$ theory.  The last expression can be expanded perturbatively around the Gaussian theory (See for e.g.   \cite{kardar2007statistical}) as follows  
\begin{align}
\ln\expval{\exp^{-\mathcal{K}}}_m = - \expval{\mathcal{K}}_m + \frac{1}{2}\left(\expval{\mathcal{K}^2}_m - \expval{\mathcal{K}}^2_m\right) + ...
\end{align}
We  keep upto only the $2^{nd}$ order in $\mathcal{K}$, as this will introduce corrections of  $\order{d^4}$ to the superfluid free energy density, which will be responsible for stabilizing the A state. The first term  $\expval{\mathcal{K}}_m $ can be evaluated and will introduce a correction to quadratic term in the superfluid Hamiltonian
\begin{align}
-\expval{\mathcal{K}}_m =& - \frac{\left(3K_1+K_2\right)}{2}\int \frac{d^3\mathbf{q}_1d^3\mathbf{q}_2d^3\mathbf{q}_3d^3\mathbf{q}_4}{(2\pi)^{12}} (2\pi)^3 \delta^3(\mathbf{q_1}+\mathbf{q_2}+\mathbf{q_3}+\mathbf{q_4}) \frac{(2\pi)^3\delta^3(\mathbf{q_1}+\mathbf{q_2})}{A_1+ B q_1^2}\tilde{d}_{i\alpha}(\mathbf{q_3})\tilde{d}^*_{i\alpha}(\mathbf{q_4}).\nonumber\\
=&- \frac{\left(3K_1+K_2\right)}{2}\int \frac{d^3\mathbf{q}_3}{(2\pi)^3}\tilde{d}_{i\alpha}(\mathbf{q_3})\tilde{d}^*_{i\alpha}(\mathbf{-q_3})\int \frac{d^3\mathbf{k}}{(2\pi)^3}\frac{1}{A_1+ B k^2},
\end{align}
  Performing an inverse Fourier transformation we  obtain the  following correction to the bare  superfluid free energy density 

\begin{align}
\beta f_{eff}&=\left(\alpha +\frac{\left(3K_1+K_2\right)(2\pi)^3}{2\Lambda^3}\int^{\Lambda^3} \frac{d^3\mathbf{k}}{(2\pi)^3}\frac{1}{A_1+ B k^2} \right)d_{i\alpha}d^*_{i\alpha}+\beta_1 d_{i\alpha}d_{i\alpha}d^*_{j\beta}d^*_{j\beta}+\beta_2d_{i\alpha}d_{j\alpha}d^*_{i\beta}d^*_{j\beta}\nonumber\\
&+\beta_3d_{j\alpha}d_{j\beta}d^*_{i\alpha}d^*_{i\beta}
+\beta_4 d_{i\alpha}d^*_{i\alpha}d_{j\beta}d^*_{j\beta}+\beta_5d^*_{i\alpha}d_{j\alpha}d_{i\beta}d^*_{j\beta}, 
\end{align} 
where $\Lambda^3$ is a cutoff wavelength. We see that first order correction from  spin fluctuation feedback effect  has changed the bare $T_c$. The coefficient of the quadratic term is  the inverse susceptibility i.e. $\chi^{-1} = A_1 +Bq^2 +\frac{\left(3K_1+K_2\right)(2\pi)^3}{\Lambda^3}\int^{\Lambda^3} \frac{d^3\mathbf{k}}{(2\pi)^3}\frac{1}{A_1+ B k^2}  $.    Next we  calculate the crucial second order $\frac{1}{2}\left(\expval{\mathcal{K}^2}_m - \expval{\mathcal{K}}^2_m\right)$ correction which will change the coefficient of the quartic terms in the bare free energy density. We follows the exact same process as outlined above and after inverse Fourier transforming we get the following free energy density
 
\begin{align}\label{eff1_He}
&\beta f_{eff}=\left(\alpha +\frac{\left(3K_1+K_2\right)(2\pi)^3}
{2\Lambda^3}\int^{\Lambda^3} \frac{d^3\mathbf{k}}{(2\pi)^3}\frac{1}{A_1+ B k^2} 
\right)d_{i\alpha}d^*_{i\alpha}+\beta_1 d_{i\alpha}d_{i\alpha}d^*_{j\beta}d^*_{j
\beta}+\beta_2d_{i\alpha}d_{j\alpha}d^*_{i\beta}d^*_{j\beta}\nonumber\\
&+\left(\beta_3-\frac{K^2_2(2\pi)^3}{4\Lambda^3}\int^{\Lambda^3} 
\frac{d^3\mathbf{k}}{(2\pi)^3}\frac{1}{(A_1+ B k^2)^2}\right)d_{j\alpha}d_{j\beta}
d^*_{i\alpha}d^*_{i\beta}+\left(\beta_5-\frac{K^2_2(2\pi)^3}{4\Lambda^3}
\int^{\Lambda^3} \frac{d^3\mathbf{k}}{(2\pi)^3}\frac{1}{(A_1+ B k^2)^2}\right)d^*_{i
\alpha}d_{j\alpha}d_{i\beta}d^*_{j\beta}\nonumber\\
&+\left(\beta_4-\frac{\left(K^2_2+6K_1^2+4K_1K
_2\right)(2\pi)^3}{2\Lambda^3}\int^{\Lambda^3} \frac{d^3\mathbf{k}}{(2\pi)^3}\frac{1}
{(A_1+ B k^2)^2}\right) d_{i\alpha}d^*_{i\alpha}d_{j\beta}d^*_{j\beta}\nonumber\\
&-\frac{K^2_3(2\pi)^3}{2\Lambda^3}\int^{\Lambda^3} \frac{d^3\mathbf{k}}
{(2\pi)^3}\frac{1}{(A_1+ B k^2)}
\left(i\vec{d}\times \vec{d}^*\right)^2, 
\end{align}

Let us now consider a simpler situation, where we ignore the cost of spatial variation in the spin fluctuation order parameter (i.e. $B$ = 0). Thus we have the following coupling
\begin{align}
\beta f_{sf-m}= A_1 m_im_i+K_1 
m_im_id_{j\alpha}d^*_{j\alpha}+K_2m_im_jd_{i\alpha}d^*_{j\alpha} +K_3 im_i(\epsilon_{ijk}d_{j\alpha}d_{k\alpha}^*),
\end{align}

The magnetic partition function is given as 
\begin{align}
\mathcal{Z}_m= \int  Dm_ie^{- \beta \mathcal{H}} =  \int  Dm_ie^{-\int d^3x  f_{sf-m}} =   \int  Dm_ie^{-\int d^3x  A_{ij}m_im_j},
\end{align}
where $A_{ij}$ contains couplings between magnetic and superconducting orders, and give corrections the bare superconducting free energy density.  Integrating out the quadratic (Gaussian) magnetic fluctuations, gives an effective free energy density

 \begin{align}\label{eff2_He}
\beta f_{eff} &= \left(\alpha+\frac{3K_1}{2A_1}+\frac{K_2}{2A_1}\right)d_{i\alpha}d^*_{i\alpha}+\beta_1 d_{i\alpha}d_{i\alpha}d^*_{j\beta}d^*_{j\beta}+\beta_2d_{i\alpha}d_{j\alpha}d^*_{i\beta}d^*_{j\beta}+\left(\beta_3-\frac{K_2^2}{4A_1^2}\right)d_{j\alpha}d_{j\beta}d^*_{i\alpha}d^*_{i\beta}\nonumber\\
&+\left(\beta_4-\frac{K_2^2+6K_1^2+4K_2K_1}{2A_1^2}\right)d_{i\alpha}d^*_{i\alpha}d_{j\beta}d^*_{j\beta}+\left(\beta_5-\frac{K_2^2}{4A_1^2}\right)d^*_{i\alpha}d_{j\alpha}d_{i\beta}d^*_{j\beta}-\frac{K_3^2}{2A_1}\left(i\vec{d}\times \vec{d}^*\right)^2,
	\end{align} 

Now if we compare Eq. \eqref{eff1_He} and Eq. \eqref{eff2_He}, we note that even though the corrections introduced to the free energy density have different forms in both cases, the sign of the correction introduced is the same, i.e. in both cases the  spin fluctuation feedback effect   leads to corrections which can stabilize a different state than the state preferred by the weak coupling theory. Subsequently, for clarity, we use the simplified result since it yields qualitatively the same results.

  The  $\beta_i$ quartic terms in Eq. \eqref{eff2_He}  are  associated with the free energy density without coupling to fluctuations, here assumed to be derived from weak-coupling theory and the quartic terms with the $K$'s originate from  spin fluctuation feedback effect. 

For large paramagnetic fluctuations, i.e. a small but positive  $A_1$, the terms that dominate are those that are proportional to $A_1^{-2}$ and we ignore terms of order $A_1^{-1}$ (for e.g. couplings of the form $m^2d^4$ ). The $K_3$ term in Eq. \eqref{eff2_He} shows that paramagnetic fluctuations can favor non-unitary states \cite{gorkov1987superconductivity,mineev1999introduction,choi1989ch,walker2002model},  but will neglected here  due to it's weaker $A_1^{-1}$ dependence. 

Weak coupling theory gives $\beta_2=\beta_3=\beta_4=-\beta_5=2\beta_1=\frac{6}{5}s$, 
where $s$ is 
a positive valued constant \cite{leggett1975theoretical,vollhardt2013superfluid}.
 When  the  spin fluctuation feedback effect  is turned off (i.e. $K_i=0$), the BW state is 
energetically  favorable with $f_{eff} =  \frac{5}{3}s$, while the A state has a 
slightly larger free energy density of $f_{eff} = 2s$. The  spin fluctuation feedback effect  coupling 
 lowers the energy of the A state by $K_2^2/3A_1^2$  compared to 
the B state and thus for large fluctuations (i.e. $K_2^2/3A_1^2 > \frac{1}{3}s$) can 
stabilize the A state. 
\cite{kuroda_paramagnon_1975}

The  A--B transition stems from the 
different temperature dependence of the weak coupling quartic terms versus the  spin fluctuation feedback effect corrections to the quartic terms.  
Quartic (i.e. $\beta$) terms originating from weak coupling theory generically have a $\frac{1}{\text{T}^2}$ dependence \cite{mineev1999introduction}, while microscopic calculations  show that 
 quartic terms  originating from the  spin fluctuation feedback effect  have a $\frac{1}{\text{T}}$ dependence  \cite{brinkman1974spin,leggett1975theoretical,kuroda_paramagnon_1975}. These calculations assumed quadratic, Gaussian spin fluctuations and imply that at high temperatures, strong fluctuations may stabilize the 
A phase, while at lower temperature  the weak coupling terms will dominate and the system 
will undergo a first order transition into the state preferred by weak coupling theory i.e. the B phase.  
Recent  calculations  done for twisted  bilayer graphene \cite{kozii2018nematic},  show that  the same temperature dependence of the quartic terms  is seen for both spin density and charge density wave fluctuations, again considering Gaussian fluctuations. This suggests that the weak-coupling quartic terms and spin-fluctuation corrections to these terms generically have a $1/T$ and $1/T^2$ temperature dependence respectively.  We will assume this to be the case in the correlated fermion materials considered below.

\label{UPt3}
\section{$\upt$}

$\upt$ is a hexagonal crystal with  $D_{6h}$   point group symmetry and has
two distinct phases under zero field, a high temperature  A phase and a low temperature B phase \cite{stewart1984possibility,joynt2002superconducting}. However, unlike $\He$, the 
A phase is time-reversal symmetric while the B phase is  time-reversal symmetry-breaking  as seen in muon spin relaxation ($\mu$SR) 
and polar Kerr measurements \cite{luke1993muon,schemm2014observation}.   $\upt$  has four 2D (reps) labeled $E_{1u/g}$ and $E_{2u/g}$, where the order parameter transforms like $\eta_1 \sim k_xk_z$, $\eta_2 \sim k_yk_z$ and $\eta_1 \sim k_x^2-k_y^2$, $\eta_2 \sim 2k_xk_y$ respectively.   The free energy density is the same for all the $E$ reps  and is given as \cite{sigrist1991phenomenological}.
\begin{align}
\beta f_{sc} 
= \alpha\left(\abs{\eta_1}^2+\abs{\eta_1}^2\right)+\beta_1\left(\abs{\eta_1}^2+\abs{\eta_1}^2\right)^2+ \beta_2\abs{\eta_1^2+\eta_2^2}^2
\end{align} 
 The coefficient of  $\abs{\eta_i^2}^2$ determines the behavior below $T_c$, with $\beta_2 > 0$ favoring the time-reversal symmetry-breaking $(1,i)$ state, 
and $\beta_2 < 0$ stabilizing the time-reversal symmetric $(1,0)$ state. To explain the existence of 
multiple phases, the currently accepted model involves coupling superconductivity  to  antiferromagnetic order
\cite{machida1989unconventional, hess1989broken}. This splits the $T_c$ between $\eta_1$ and $\eta_2$ and allows for two transitions.
 
However, recent experiments raise questions 
over the existence of true antiferromagnetic order near the two closely spaced superconducting transitions. The Bragg peaks in inelastic nuetron scattering are not resolution limited near the  superconducting transitions of $ T_{cA}= 530mK$ and $T_{cB} =  480 mK$ \cite{joynt2002superconducting,koike1998long,koike1999neutron}, in addition to the absence of signatures of a magnetic transition in specific heat, magnetization and NMR Knight shift experiments around this temperature.     The peaks   start narrowing only below 50mK and become resolution limited at 20 mK \cite{koike1998long,koike1999neutron}, which seems consistent with anomalies seen in specific heat \cite{schuberth1992specific,schuberth1996some}, thermal expansion \cite{sawada1996thermal} and magnetization measurements \cite{schottl1999anisotropic} seen near 20mK. Interestingly, NMR experiments show an anomaly at 50mK which is associated with the fluctuations slowing down, though there is no sign of static order down till 15mK \cite{kitaoka2000nmr,van2002weak}.     

This has lead to the current  interpretation that  these 
experiments  imply the presence of antiferromagnetic fluctuations \cite{gannon2017spin},  
instead of  antiferromagnetic order, and we suggest   that a  generalized  spin fluctuation feedback effect  
then  stabilizes a time-reversal symmetric A state.

Our theory is a Ginzburg-Landau theory and is valid 
near the A--B transition and cannot be extended deep into the B state.  However, the use of  spin fluctuation feedback effect  
fluctuations to generate two closely spaced transitions should be valid near the 
superconducting transitions (i.e. $T_c$ $\sim$ 500mK).  We note that the Ginzburg temperature for magnetic transitions is generically of the order $1-10^{-2}$mK of the magnetic transition temperature ( which for $\upt$ is  $T_{cm}\sim 50mK$). Consequently,  the superconducting A--B transitions will be insensitive to critical phenomena stemming from the possible ultra low temperature magnetic ordering.  Additionally, there are no signatures of  quantum critical effects in this material and the other materials considered in this manuscript.  Hence arguments similar to those developed for superfluid $\He$ can be used to explain two transitions in $\upt$, and it suffices to consider a Gaussian theory of spin fluctuations.  

We shall proceed  analogously to $\He$, and assume that the B time-reversal symmetry-breaking  state is favored by  the weak coupling theory, while strong fluctuations can stabilize a time-reversal symmetric A phase.  The fluctuations 
are characterized by  wave vectors \textbf{Q}$_1$=$\frac{1}{2}$
 \textbf{a}$^*$, \textbf{Q}$_2$=$\frac{1}{2}$(\textbf{b}$^* $-\textbf{a}$^*$), \textbf{Q}$_3$= -$
 \frac{1}{2}$\textbf{b}$^*$ \cite{1988PhRvL..60..615A,koike1998long}  which are 
 associated with the magnetic order  parameters $m_1$, $m_2$ and  $m_3$   
 respectively. The coupling of superconductivity to the magnetic 
 fluctuations  is constructed to be invariant under $D_{6h} \times U(1)\times \mathscr{T}$, where $\mathscr{T}$ is time-reversal symmetry and is expressed as \cite{machida1989unconventional}
\begin{align}
\beta f_{sc-m} =A_1m_i^2+ K_1(m_i^2)(\eta_j\eta_j^*)+ K_2 \left[(2m_1^2 - m_2^2-m_3^2)(\abs{\eta_1}^2-\abs{\eta_2}^2)+\sqrt{3}(m_3^2-m_2^2)(\eta_1\eta_2^*+\eta_2\eta_1^*)\right] .
\label{coup}
\end{align}
 Integrating out the fluctuations as before gives the following  free energy density: 
	\begin{align}\label{eff1_upt}
	\left(\alpha + \frac{3}{2}\frac{K_1}{A_1}\right)\left(\abs{\eta_1}^2+\abs{\eta_1}^2\right)+\left(\beta_1-\frac{3}{4}\frac{K_1^2}{A_1^2}\right)\left(\abs{\eta_1}^2+\abs{\eta_1}^2\right)^2+\left(\beta_2-\frac{6}{4}\frac{K_2^2}{A_1^2}\right)	\abs{\eta_1^2+\eta_2^2}^2.
	\end{align}

We see that the terms originating from  the generalized  spin fluctuation feedback effect  have introduced negative valued  corrections to the quartic terms, which most importantly  changes the $\abs{\eta_i^2}^2$ coefficient. Thus for large fluctuations, these  spin fluctuation feedback effect  terms can stabilize the A state, instead of the time-reversal symmetry-breaking B by making this coefficient negative.

 This  also interestingly  implies that two transitions will occur only if the B state is a broken time-reversal symmetry  state. In particular, if the B state was time-reversal symmetric  then the  spin fluctuation feedback effect  terms would simply further stabilize the time-reversal symmetric nematic state.  

As argued earlier, the quartic terms that stem from weak-coupling theory should increase more strongly as temperature is decreased than the quartic terms which arise from the generalized spin fluctuation feedback effect. This allows the coefficient of the $\abs{\eta_1^2+\eta_2^2}^2$ to change sign as temperature is decreased,  so that a transition into the a broken time-reversal symmetry state is possible. We discuss this in more detail below.

\subsection{Effective theory for A--B transitions} 
A complete phenomenological description of a second phase transition within a single multi-dimensional rep requires a free energy density that is at least  eighth order in the order parameter \cite{1995JETP...80..485G,toledano1987landau}. For this reason, we consider a simpler approach and model the A--B transition as an effective phenomenological  theory in which we start with  $(1,0)$ state for $\upt$ and allow the B state to continuously grow out of this  i.e. $(1+\tilde{\eta}_{1i}, 0+\tilde{\eta}_{2i})$, where $\tilde{\eta}_{i}$ is small near the  transition.  Time-reversal symmetry  allows us to 
classify the order parameter  $\tilde{\eta}_{i}$ for the A--B transition into a real part $\tilde{\eta}_R$ which is invariant under $\mathscr{T}$, and an imaginary part $\tilde{\eta}_I$ which changes sign under $\mathscr{T}$, with the following transformation properties.

\begin{align}
 \tilde{\eta}_R  \overset{\mathcal{T}} \rightarrow  \tilde{\eta}_R  \hspace{3.5em} \tilde{\eta}_I \overset{\mathcal{T}} \rightarrow - \tilde{\eta}_I.
\end{align}

The condition that the second transition is observed to break time-reversal symmetry, allows us to consider only the imaginary order parameter. The  $(1,0)$ state has $D_2(C_2)\times \mathscr{T}$ and $D_2\times \mathscr{T}$  \cite{gorkov1987superconductivity} symmetry for the 
$E_{1u/g}$ and $E_{2u/g}$ reps respectively.  The order parameter $\tilde{\eta}_{1I}$ belongs 
to the $A_1$ rep of $D_2$, while $\tilde{\eta}_{2I}$ belongs to the $B_1$ 
rep.  Hence our mechanism only allows these two possible symmetries for the $B$ phase.

\subsection{Constraints from polar Kerr effect}
The observation of a polar Kerr signal for the A--B transition \cite{schemm2014observation} further constrains the possible order parameters. In particular, polar Kerr experiments shows that signal can be trained with an applied magnetic field \cite{schemm2014observation}. This implies that the only viable order parameters are those which belong to the same representation as a component of the magnetic field   ($H_x$, $H_y$, $H_z$). This follows because a Kerr signal that can be trained in field is only possible if the superconducting order parameter couples linearly to the applied field.  These order parameters   shall be referred to as Kerr active (in the literature, this is sometimes labeled  as belonging to a ferromagnetic class
\cite{volovik1985superconducting}). This rules out the $A_1$ order parameter since it is 
Kerr inactive. The $\tilde{\eta}_{2I}$ order parameter has $H_z$ symmetry, which is consistent with the direction of the training field applied along the c-axis in the experiment.   Thus the  A--B transition can be modeled by the effective order 
parameter $\tilde{\eta}_{2I}$ with the following free energy $\beta f_{A\rightarrow B} = \alpha_{1I} 
\tilde{\eta}
_{2I}^2 +\beta_{1I}\tilde{\eta}_{2I}^4$. We shall use this approach subsequently to constrain the possible symmetries of the order parameters.

\section{$\pos$}
 $\pos$ (POS) is a Pr based tetrahedral correlated fermion skutterudite superconductor with a $T_h$ 
point group, and like $\upt$ has two distinct phases \cite{vollmer2003low}.  Polar Kerr and $\mu$SR 
measurements  show a time-reversal symmetry-breaking B phase \cite{aoki2003time,levenson2018polar}, while the 
A phase is time-reversal symmetric. POS has been studied by phenomenological methods \cite{curnoe}, 
however there is no satisfactory mechanism for the double transition. Inelastic nuetron scattering experiments 
indicate the presence of  antiferro-quadrupolar fluctuations with a $\textbf{Q}$ = (1, 0, 0)
\cite{kaneko2007neutron,kuwahara2005direct}, which is a single \textbf{Q} order, invariant under the point group operations.   The order 
parameter of these antiferro-quadrupolar fluctuations is 3D with  components that transform as $m_1 
\sim k_y k_z$, $m_2 \sim k_x k_z$ and $m_3 \sim k_x k_y$ 
\cite{shiina_multipolar_2004,shiina_theory_2004}. The antiferro-quadrupolar fluctuations can stabilize a time-reversal symmetric A phase for both the $E$ and $T$ reps as shown below. 

For the $E$ reps,  where the order parameter transforms as $\eta_1 \sim 2k_z^2-
k_x^2-k_y^2$, $\eta_2 \sim k_x^2-k_y^2$, the coupling is 
	\begin{align}\label{coupEpos}
	\beta f_{sc-m} &=A_1(m_i^2)+ K_1(m_j^2)(\eta_j\eta_j^*)
			+ K_2 \left[(2m_3^2 - m_1^2-m_2^2)(\abs{\eta_1}^2-\abs{\eta_2}^2)-\sqrt{3}(m_1^2-m_2^2)(\eta_1\eta_2^*+\eta_2\eta_1^*)\right]\nonumber\\
			&+ K_3\left[\sqrt{3}(m_1^2-m_2^2)(\abs{\eta_1}^2-\abs{\eta_2}^2)+(\eta_1\eta_2^*+\eta_2\eta_1^*)(2m_3^2 - m_1^2-m_2^2)\right].
	\end{align}
As noted in $\upt$, we shall consider the simple case of $B (\nabla \vec{m})^2=0$. Integrating out the antiferro-quadrupolar fluctuations, we obtain the effective free energy density
\begin{align}\label{posEeff}
		\beta f_{eff} = \left(\alpha+\frac{3K_1}{2A_1}\right)(\eta_j\eta_j^*)+\left(\beta_1-\frac{3K_1^2}{4A_1^2}-\frac{3K_2^2}{2A_1^2}-\frac{3K_3^2}{2A_1^2}\right)\left(\eta_j\eta_j^*\right)^2 + \left(\beta_2 -\frac{3K_2^2}{2A_1^2}-\frac{3K_3^2}{2A_1^2}\right)(\eta_1\eta_2^*-\eta_2\eta_1^*)^2.
		\end{align}
This generalized  spin fluctuation feedback effect  may again stabilize a time-reversal symmetric A state $(\phi_1,
\phi_2)$ with  $D_2 \times \mathscr{T}$ symmetry \cite{curnoe,mukherjee2006microscopic}, instead of the time-reversal symmetry-breaking B phase $(1,i)$ with 
$T(D_2)$ symmetry by changing the sign of the $(\eta_1\eta_2^*-\eta_2\eta_1^*)^2$ term from positive to negative. The A phase has the two components with the  same magnitude but an arbitrary phase \cite{curnoe}. Similar to $\upt$ two transitions are possible only when the B state is time-reversal symmetry-breaking. The A--B transition  is modeled similar to $\upt$, however both $\eta_{1I/2I}$ have $A_1$ Kerr inactive symmetry and are ruled out due to presence of Kerr effect, thereby eliminating the 2D E reps scenario for $\pos$.Thus we can see just using these general symmetry considerations we are able to rule out a pairing scenario with this mechanism.      

For the 3D $T_{g/u}$ reps, with order parameter which transforms for example as $\eta_1\sim k_yk_z$, $\eta_2\sim k_xk_z$ and $\eta_3\sim k_xk_y$, the coupling is 
\begin{align}\label{coupTpos}
	\beta f_{sc-m}& = A_1\left(m_i^2\right)+K_1(\eta_j\eta_j^*)\left(m_j^2\right)						 +K_2\left[3\left(\abs{\eta_2}^2-\abs{\eta_3}^2\right)\left(m_2^2-m_3^2\right)+\left(2\abs{\eta_1}^2-\abs{\eta_2}^2-\abs{\eta_3}^2\right)\right.\nonumber\\
	&\times\left.\left(2m_1^2-m_2^2-m_3^2\right)\right]+K_3\left[\left(\abs{\eta_2}^2-\abs{\eta_1}^2\right)		\left(2m_1^2-m_2^2-m_3^2\right)-\left(2\abs{\eta_1}^2-\abs{\eta_2}^2-\abs{\eta_3}^2\right)\left(m_2^2-m_3^2\right)\right]\nonumber\\
	&+K_4\left[\left(\eta_2\eta_3^*+\eta_2^*\eta_3\right)m_2m_3+\left(\eta_3\eta_1^*+\eta_3^*\eta_1\right)m_3m_1+\left(\eta_1\eta_2^*+\eta_1^*\eta_2\right)m_1m_2\right].
	\end{align}  
 Integrating out the Gaussian antiferro-quadrupolar fluctuations gives the following effective free energy density: 
	\begin{align}
	\beta f_{eff} &= \left(\alpha+3\frac{K_1}{A_1}\right)(\eta_j\eta_j^*)+\left(\beta_1-\frac{6K_1^2}{A_1^2}+\frac{6K_2^2}{A_1^2}+\frac{2K_3^2}{A_1^2}-\frac{K_4^2}{4A_1^2}\right) \left(\eta_j\eta_j^*\right)^2+\left(\beta_2-\frac{K_4^2}{4A_1^2}\right)\abs{\eta_i^2}^2\nonumber\\
	&+\left(\beta_3-\frac{18K_2^2}{A_1^2}-\frac{6K_3^2}{A_1^2}+\frac{K_4^2}{2A_1^2}\right) \left(\abs{\eta_1}^4+\abs{\eta_2}^4+\abs{\eta_3}^4\right).
	\end{align}

 Again the generalized  spin fluctuation feedback effect  has changed the coefficient of the  bare  free energy density and hence allows for the possibility  of a time-reversal symmetric A state. Interestingly, here we may have two transition even if the A state is time-reversal symmetry-breaking, due the indeterminant sign of the  correction to  the $\beta_3$ coefficient. However since this does not agree with the experimental identification of the B state having broken time-reversal symmetry,  we don't  consider this possibility. This rep has two states 
which are time-reversal symmetric, the $(1,0,0)$ state with $D_2(C_2)\times \mathscr{T}$ symmetry  and the 
$(1,1,1)$ state with $C_3\times \mathscr{T}$ symmetry. Both of these allow for a 
transition to a time-reversal symmetry-breaking B state which is Kerr active and hence provide two viable channels for the transition. The physics of this is similar to the 2D rep case for $\upt$ and  is worked out in the appendix, the results of which are collected in Table \ref{tab:table1}. 
It would be of interest to carry out Kerr measurements \cite{levenson2018polar} under different directions of training field to further constrain the possible pairing channels.

\section{$\ube$}

 $\ube$ is a cubic material with an $\text{O}_h$ point group, which also has  
 two transitions \cite{heffner1990new,zieve2004pressure}, but only  for a doping range of 2 \% $<$ x $<$ 4 \%.  The B phase is again a time-reversal symmetry-breaking state \cite{heffner1990new}. Antiferromagnetic fluctuations are seen in inelastic nuetron scattering  with a wave vector of \textbf{Q$_3$} = (1/2, 1/2, 0) \cite{hiess2002neutron}.  We  consider both the $E$ and $T$ reps and model the system with O$_h$ symmetry, having antiferromagnetic fluctuations with wave vector \textbf{Q$_3$}. The star of  \textbf{Q$_3$} gives two additional wave vectors \textbf{Q$_2$} = (1/2, 0, 1/2) and \textbf{Q$_1$} = (0, 1/2, 1/2), each of which correspond  to a 1D order parameter $m_1$, $m_2$, $m_3$. Here $\eta_1$ and $\eta_2$ transform exactly as $E$ reps of $\pos$. The coupling  is   
		\begin{align}
	\beta f_{sc-m} &=A_1(m_i^2)+ K_1(m_i^2)(\eta_j\eta_j^*)
			+ K_2 \left[(2m_3^2 - m_1^2-m_2^2)(\abs{\eta_1}^2-\abs{\eta_2}^2)-\sqrt{3}(m_1^2-m_2^2)(\eta_1\eta_2^*+\eta_2\eta_1^*)\right].
	\end{align}
	 The $K_3$ term present in Eq. \ref{coupEpos} is absent above, due to additional  symmetry elements present in the $O_h$ as compared to $T_h$ point group.   The correction to the free energy density from these fluctuations is 
 	
 	\begin{align}
		\beta f_{eff} = \left(\alpha+\frac{3K_1}{2A_1}\right)(\eta_j\eta_j^*)+\left(\beta_1-\frac{3K_1^2}{4A_1^2}-\frac{3K_2^2}{2A_1^2}\right)\left(\eta_j\eta_j^*\right)^2 + \left(\beta_2 -\frac{3K_2^2}{2A_1^2}\right)(\eta_1\eta_2^*-\eta_2\eta_1^*)^2 .
		\end{align}
There are two possible  time-reversal symmetric A states. The details of the A--B effective theory follows that of $\upt$ and  is in the appendix, with  the  possible A state being  (1,0) with $D_4\times \mathscr{T}$ symmetry and (0,1) 
with $D_4^{(1)}(D_2)\times\mathscr{T}$ symmetry \cite{volovik1985superconducting}.    These states are  Kerr inactive, and we suggest that a polar Kerr measurement be performed on this material. \textit{If a field-trainable polar Kerr signal is seen then the 2D order parameter can be ruled out.}
 
 For the 3D order parameter case, we note that, $\ube$ has four $T$ reps, one which transforms exactly like $T$ reps of $\pos$ which we call $T_{2g/u}$ and the other $T_{1g/u}$ which transforms as $\eta_1\sim k_yk_z(k_y^2-k_z^2)$, $\eta_1\sim k_zk_x(k_z^2-k_x^2)$ and $\eta_3\sim k_xk_y(k_x^2-k_y^2)$.  The coupling for  these  $T$ reps is 
\begin{align}
	 f_{sc-m} = A_1\left(m_i^2\right)+K_1\left(m_i^2\right)(\eta_j\eta_j^*)						 +K_2\left[3\left(\abs{\eta_2}^2-\abs{\eta_3}^2\right)\left(m_2^2-m_3^2\right)+\left(2\abs{\eta_1} ^2-\abs{\eta_2}^2-\abs{\eta_3}^2\right)\left(2m_1^2-m_2^2-m_3^2\right)\right].
	\end{align}
 The $K_3$ term found in Eq. \ref{coupTpos} is absent because of higher $O_h$ symmetry compared to the $T_h$ symmetry, while the $K_4$ term is forbidden as $m_im_j$ is not translationally invariant for this \textbf{Q} vector. The effective free energy density  obtained is  
\begin{align}\label{effTube}
	\beta f_{eff} = \left(\alpha+3\frac{K_1}{A_1}\right)(\eta_j\eta_j^*)+\left(\beta_1-\frac{3K_1^2}{4A_1^2}+\frac{3K_2^2}{A_1^2}\right)\left(\eta_j\eta_j^*\right)^2+\beta_2\abs{\eta_j^2}^2
	+ \left(\beta_3-\frac{9K_2^2}{A_1^2}\right) \left(\abs{\eta_1}^4+\abs{\eta_2}^4+\abs{\eta_3}^4\right). 
	\end{align}
	
	Interestingly, here unlike $\pos$ there is no change to $\beta_2$, which is a result of the $K_4$ coupling being absent for Eq. \ref{effTube} in  contrast to Eq. \ref{coupTpos}.  There are   four  possible time-reversal symmetric A state, two for each $T$ reps, i.e. (1,0,0) and (1,1,1). However due to there being no change to coefficient of  the $\abs{\eta_i^2}^2$ term, the  (1,0,0) is the only  time-reversal symmetric A state which allows for a viable transition to a time-reversal symmetry-breaking B state \cite{gorkov1987superconductivity}.  This state has $D_4(C_4)\times\mathscr{T}$ and 
$D_4^{(2)}(D_2)\times\mathscr{T}$  symmetry 
for the $T_1$ and $T_2$ reps respectively \cite{volovik1985superconducting}. The A--B transition, follows similar to $\upt$, and is modeled in the appendix. 
\textit{Polar Kerr measurements especially if trainable by field,  may be useful as  they  can rule out the $E_{g/u}$ reps scenario, and can eliminate  other 3D $T_{g/u}$ order parameters.}

\section{Conclusions}
In analogy to superfluid $\He$, we have argued that a generalized  spin fluctuation feedback effect  can account for multiple transitions seen in correlated fermion materials. We have  provided a simple phenomenological framework to capture this and show that this naturally  provides a unifying mechanism for two superconducting transitions observed in $\upt$, $\pos$ and $\ube$.  In addition, the use of this generalized  spin fluctuation feedback effect allows us to constrain various pairing symmetries. The results of this analysis  are tabulated in Table  \ref{tab:table1}. In particular, we are able to rule out the 2D ($E_{g/u}$) scenario for $\pos$, while  for the  the 3D ($T_{g/u}$) reps  of $\ube$ only one of two possible  A states is allowed. Additionally,  if a polar Kerr signal is observed for $\ube$, we will be able to rule out the 2D ($E_{g/u}$) for $\ube$, and place further constraints on other pairing channels.

\textit{Acknowledgments}.- Adil Amin and D.F.A. acknowledge
financial support through the UWM research growth initiative. We would like to thank Aharon Kapitulnik, Bill Halperin,  Deep Chatterjee and Joseph O' Halloran for useful discussions.

\appendix

\section{Effective  theory for A--B transition}

\textit{\pos}.- 
For the  3D $T_{1g/u}$ reps, we have two possible time-reversal symmetric A states the $(1,0,0)$ state with $D_2(C_2)\times \mathscr{T}$ symmetry  and the 
$(1,1,1)$ state with $C_3\times \mathscr{T}$ symmetry. Both of  these allow for a 
transition to a time-reversal symmetry-breaking B state which is Kerr active and hence provide  two viable channels for the transition. In the case of $(1+\tilde{\eta}_{1I}, 
0+\tilde{\eta}_{2I}, 0+\tilde{\eta}_{3I})$, 
$\tilde{\eta}_{1I}$ belongs to the $A_1$  Kerr inactive reps and  hence is not 
considered, while $\tilde{\eta}_{2I}$ and $\tilde{\eta}_{3I}$ belong to the 
 Kerr active $B_1$ and $B_2$ reps respectively. The form of the free 
energy will look similar to the 1D  A--B transition in $\upt$, and the exact 
transition would depend on which rep has a higher $T_c$. The free energy will be \begin{align}
\beta f_{A\rightarrow B} = \alpha_{I} \tilde{\eta}
_{2/3I}^2 +\beta_{I}\tilde{\eta}_{2/3I}^4.
\end{align}  
 
  For the $(1+\tilde{\eta}_{1I}, 1+\tilde{\eta}_{2I}, 
1+\tilde{\eta}_{3I})$ state we have a 1D $A_1$ 
 Kerr inactive order parameter defined as 
$\eta_A=\tilde{\eta}_1+\tilde{\eta}_2+\tilde{\eta}_3$ and which  is ignored, and a 2D Kerr
active $E$ order parameter, defined as  
\begin{align}
\eta_x=\frac{1}{\sqrt{6}}(2\tilde{\eta_3}-\tilde{\eta_2}-\tilde{\eta_3}), \hspace{0.5em} \eta_y
=\frac{1}{\sqrt{2}}(\tilde{\eta_1}-\tilde{\eta_2}),
\end{align}
 where $\eta_x,\eta_y$ belong to the 2D Kerr active $E$ reps of $C_3$. The free energy for this $E$ rep is 
 \begin{align}
\beta f^E_{A\rightarrow B}=\alpha_I \eta_{i}^2 +\beta_{1I}\left(\eta_{i}^2\right)^2+\gamma\left(\eta_{i}^2\right)^3+\gamma_1\eta_{xI}^2(\eta_{xI}^2-3\eta_{yI}^2)^2+\gamma_2\eta_{yI}^2(3\eta_{xI}^2-\eta_{yI}^2)^2
+\gamma_3\eta_{xI}\eta_{yI}(\eta_{xI}^2-3\eta_{yI}^2)(3\eta_{xI}^2-\eta_{yI}^2)
 \end{align}
Once $\alpha_I$ changes sign the ground state will be $\eta$ = $\eta_I$($\cos\theta,\sin\theta$), where the value of $\theta$ will depend on the value of the coefficients ($\gamma_i$) of the 6$^{\text{th}}$ order terms, and hence provides a viable mechanism  for a transitions to a time-reversal symmetry-breaking B phase.

\textit{\ube}.- 
For the 2D $E_{g/u}$ reps  there are two possible  time-reversal symmetric A states. We model the A--B transition as done before, with  the   A state being  (1,0) possessing $D_4\times \mathscr{T}$ symmetry and (0,1) 
with $D_4^{(1)}(D_2)\times\mathscr{T}$ symmetry \cite{volovik1985superconducting}.  The B 
state grows as  $(1+\tilde{\eta}_{1I}, 0+\tilde{\eta}_{2I})$, where $\tilde{\eta}_{1I}$ belongs to the $A_1$  
 and $\tilde{\eta}_{2I}$ belongs to $B_1$ reps of $D_4$, while for the $(0,1)$ state, the 
situation is reversed. In that case the the B state grows as $(0+\tilde{\eta}_{1I}, 1+\tilde{\eta}_{2I})$, with with  $\tilde{\eta}_{1I}$  having  $B_1$ symmetry   
 and $\tilde{\eta}_{2I}$ belonging to the  $A_1$ rep of $D_4$.  The free energy will be the same as  for the 1D order parameters
\begin{align}
\beta f_{A\rightarrow B} = \alpha_{I} \tilde{\eta}
_{1/2I}^2 +\beta_{I}\tilde{\eta}_{1/2I}^4.
\end{align}
Interestingly due the absence of a Kerr measurement experiment, the  $A_1$ rep is viable, and  the same  physics applies here  as discussed for  the  $s+is$ states  in the iron based superconductors \cite{mettout1996unconventional,carlstrom2011length,maiti2013s+,garaud2018properties}.  These states are  Kerr inactive, and can  be ruled out depending on the results of a trainable polar Kerr measurement in $\ube$.

For the 3D reps $T_{1g/u}$ and $T_{2u/g}$ there are  four  possible time-reversal symmetric A states, two for each $T$ reps, i.e. (1,0,0) and (1,1,1). However, as explained in the main text, the  (1,0,0) is the only viable time-reversal symmetric A state which allows for a transition to a time-reversal symmetry-breaking B state, due to the form of the coupling  between superconductivity and antiferromagnetic fluctuations. This state has $D_4(C_4)\times\mathscr{T}$ and 
$D_4^{(2)}(D_2)\times\mathscr{T}$  symmetry 
for the $T_1$ and $T_2$ reps respectively \cite{volovik1985superconducting}. The A--B transition is modeled similar to  $\pos$ i.e. $(1+\tilde{\eta}_{1I}, 
0+\tilde{\eta}_{2I}, 0+\tilde{\eta}_{3I})$,  except here $\tilde{\eta}_{1I}$ belongs to the $A_1$ rep  
of $D_4$  and would again have the $s+is$ physics with the standard free energy for 1D reps
\begin{align}
\beta f_{A\rightarrow B} = \alpha_{I} \tilde{\eta}
_{1I}^2 +\beta_{I}\tilde{\eta}_{2I}^4,
\end{align} 
while $(\tilde{\eta}_{2I},\tilde{\eta}_{3I})$ belong to the 2D Kerr active $E$ reps of $D_4$.  The free energy for the $E$ rep is 
\begin{align}
\beta f^E_{A\rightarrow B}=\alpha_I \tilde{\eta}_{i} 
\tilde{\eta}_i+\beta_{1I}(\tilde{\eta}_{i} 
\tilde{\eta}_i)^2+\beta_{2I}\tilde{\eta}_{2I}^2 \tilde{\eta}^2_{3I}.
\end{align}
 For $\beta_2>0$ will pick the $\eta_I$(1,0)  and for $\beta_2<0$ the $\eta_I$(1,1) 
state, both of which break time-reversal symmetry.
\bibliography{SFFF_paper}
\end{document}